\documentclass[12pt,preprint]{aastex}
\def\paren#1{\left( #1 \right)}
\def\angle#1{\left\langle #1 \right\rangle}
\begin{document}
\title{Luminosity and Variability of Collimated Gamma-ray Bursts}
\author{Shiho Kobayashi$^1$, Felix Ryde$^2$ and Andrew MacFadyen$^3$}
\affil{
$^{1}$Department of Earth and Space Science, Osaka University,
Toyonaka, Osaka 560, Japan\newline
$^{2}$Center for Space Science and Astrophysics, Stanford University,
Stanford, CA 94305, USA\newline
$^{3}$Theoretical Astrophysics 130-33, Caltech, Pasadena, CA 91125, 
USA\newline}
\begin{abstract}
 Within the framework of the internal shock model, we study the
 luminosity and the variability in gamma-ray bursts from collimated
 fireballs. In particular we pay attention to the role of the
 photosphere due to $e^\pm$ pairs produced by the internal shock
 synchrotron photons. It is shown that the observed Cepheid-like
 relationship between the luminosity and the variability can be
 interpreted as a correlation between the opening angle of the fireball
 jet and the mass included at the explosion with a standard energy
 output. We also show that such a correlation can be a natural
 consequence of the collapsar model.  Narrow jets, in which the typical
 Lorentz factors are higher than in wide jets, can produce more variable
 temporal profiles due to smaller angular spreading time scales at the
 photosphere radius. Using a multiple-shell model, we numerically
 calculate the temporal profiles of gamma-ray bursts and show that our
 simulations reproduce the observed correlation. 
\end{abstract}
\keywords{ gamma rays: bursts;  shock waves}
\section{Introduction}
Gamma-ray bursts (GRBs) and the afterglows are well described by the
fireball model (e.g. see Piran 2000), in which an explosive flow of
relativistic matter (ejecta) is released from a central engine.
The collision of fast-moving ejecta with slower ejecta results in 
a GRB. Thereafter, the ejecta shocks and sweeps up a large amount of 
ambient matter. The shocked matter powers the long-lived afterglow.

The temporal profiles of GRBs are often very variable and each profile
looks very different. Quantitative measures of the variability have been
suggested allowing for a systematic study of their morphology. Several
studies have explored the possibility that quantities directly
measurable in GRB light curves could be related to the luminosities of
the bursts. Stern, Poutanen \& Svensson (1999) concluded that there is
an intrinsic correlation between luminosity and the complexity of
GRBs. Ramirez-Ruiz \& Fenimore (1999; see also Fenimore \& Ramirez-Ruiz
2000) found that the luminosities of seven bursts with known redshifts
are correlated with the variabilities. Based on this work, Reichart et
al. (2001) also reported a possible Cepheid-like luminosity estimator
for long bursts. Using the GRB rate density derived from this
correlation, Lloyd-Ronning, Fryer \& Ramirez-Ruiz (2001) discussed 
the star formation rate at high redshift.

\begin{figure}
\plotone{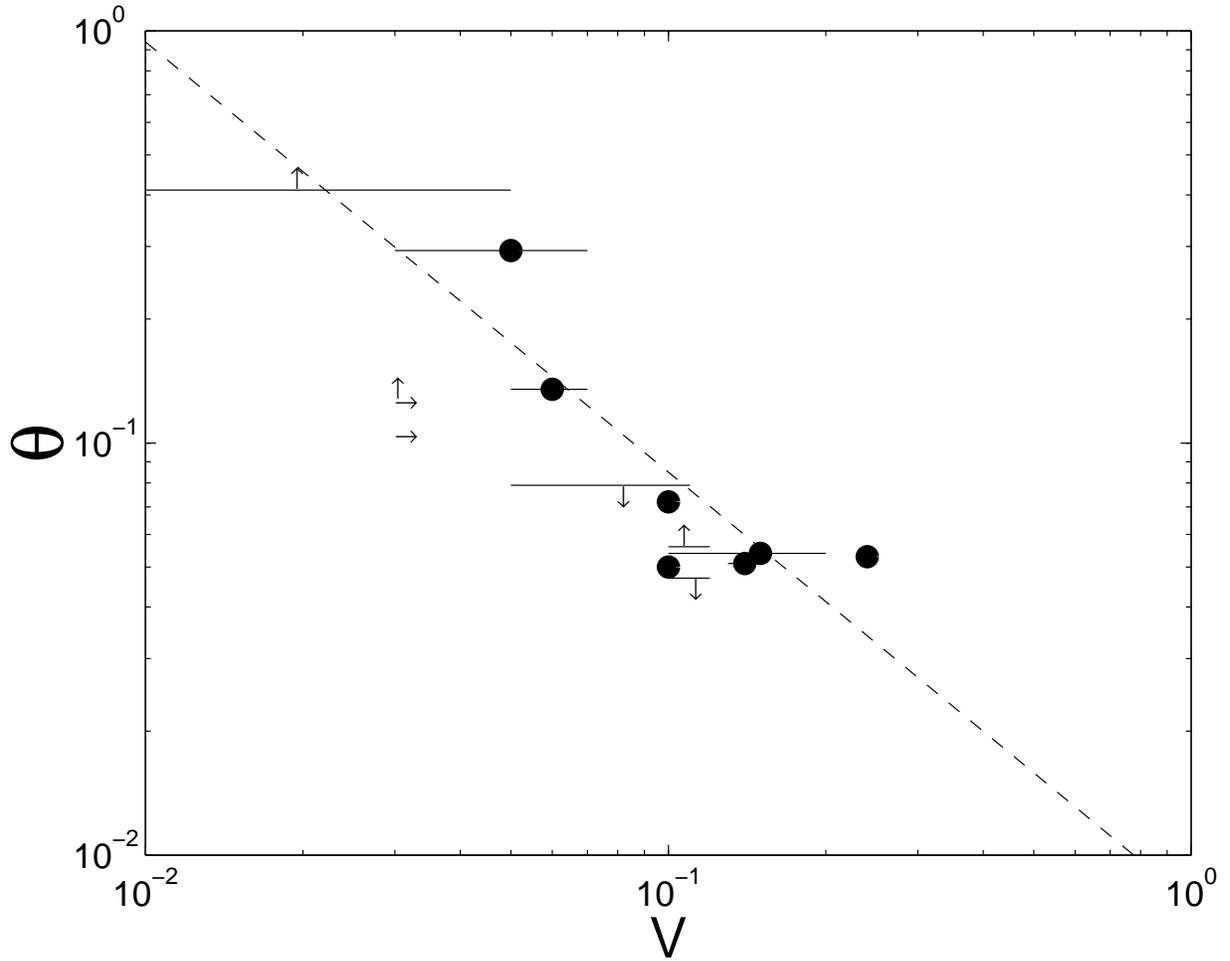}
  \caption{
  Opening Angle vs. Variability:  
  Measurements (filled circles) and upper or lower limits (arrows)
  are shown from Reichart et al (2001) and Frail et al (2001).
  A linear fit to filled circles gives $\theta \propto V^{-1.0}$.
  \label{fig:var_jet}  }
 \end{figure}

Recent afterglow observations allow us to determine the geometry of 
ejecta, whether it is spherical or conical. There is excellent
observational evidence that the ejecta have conical geometry (jet), and
there appears to be a correlation in the sense that the bursts with the
largest gamma-ray fluences have the narrowest opening angles. The
correlation is improved when the fluences are all scaled to the same
distance by using the redshift measurements. Frail et al. (2001),
Panaitescu \& Kumar (2002) and Piran et al.(2001)
reported that the gamma-ray energy releases, corrected for geometry, are
narrowly clustered around $10^{51}$erg, and suggested that the wide
variation in fluences and luminosities of GRBs is due entirely to a
distribution of the opening angles. If this is the case, the
Cepheid-like relation implies a correlation between the opening angle
and the variability: more highly collimated GRBs are more variable.
Such a correlation is actually found in the observations (see 
figure \ref{fig:var_jet}).    

As a result of relativistic beaming, an observer can see only a limited
portion of the ejecta. There should be no observable distinction between
a spherical ejecta and a conical ejecta until the ejecta has slowed down
in the afterglow phase. Therefore, the correlation between variability
and opening angle should be attributed to the properties of the central
engine. In this paper, we will explore the relations among variability,
luminosity and jet opening angle in the framework of the internal shock 
model. We show that the variability-opening angle relation and the 
Cepheid-like relation can be interpreted as a correlation between the 
opening angle and the mass included at the explosion. In \S2 we first 
discuss the collapsar model as the central engine of GRBs. In \S3 we 
make some comments on the typical GRB energy. In \S4 we discuss the 
internal shocks and the temporal profile. In \S5 we study internal
shocks by using a multiple-shell model. In \S6 we consider what
assumptions are actually require to fit the Cepheid-like relation.
In \S7 we give conclusions.      
\section{Central Engine: Collapsar}
While the nature of the GRB progenitors is still unsettled, it now
appears likely that at least some bursts originate in explosions of
very massive stars (main sequence mass $M_{ms} > 25 ~M_{\sun}$).
The model currently favoured for long bursts is the collapsar model in
which GRBs are caused by relativistic jets expelled along the rotation
axes of collapsing massive stars.  The jets are powered by a black
hole with a surrounding accreting torus.  Energy from the accretion is
pumped into jets via electrodynamic processes or by neutrino
annihilation.  In addition spin energy of the rotating black hole may
be tapped by magnetic fields anchored in the accretion disk.

According to Frail et al. (2000), Panaitescu \& Kumar (2002) the wide
variation in the fluences of GRBs originates from the differences in
opening angles of the jets. A wide jet radiates gamma-ray photons into
a large solid angle, resulting in a dim burst. Though we do not know
yet what physical mechanism results in the wide variation of the
opening angles, it is likely that a wider jet involves a larger mass 
at the explosion, and that it results in a flow with a lower Lorentz
factor.

In the collapsar model, the duration of a GRB is set by the collapse
time scale of a massive stellar core, typically $\sim 10$ sec for
helium cores of $8 < M_{\alpha} < 15$.  The progenitor
has a mass $> 25 ~M_{\sun}$ on the main sequence but loses its
envelope to a binary companion and/or stellar wind before core
collapse.  The stellar core is assumed to collapse to a black hole,
due to it's large mass, and to be rotating sufficiently rapidly to
form an accretion disk around the newly born black hole (Woosley
1993; MacFadyen \& Woosley 1999).  

Typical accretion disk time scales are milliseconds and gas typically
resides briefly in the accretion disk.  The key point is that the disk
is continuously fed by the collapsing star.  Short time scales are
available and indeed expected due both to variability in accretion rate
and the jet instabilities as the hot jet material expands along the
polar axis of the star and interacts with the surrounding star.  

For disks forming with radius $\sim 10^7$ cm, typical accretion
time scales are t$_{acc} \sim 0.01$ sec.  Fluctuations in accretion
rate due to instabilities were shown to exist with 50 msec time scales
(MacFadyen \& Woosley 1999). In models where accretion energy is tapped
to power relativistic polar jets, these fluctuations in accretion may
translate into variation in jet Lorentz factor perhaps leading to
variation in the GRB light curves.  However time scales calculated so
far in numerical simulations are probably too short to produce the
observed variations.

A more promising source of time structure in GRBs are instabilities
arising as a relativistic jet propagates through the stellar mantle.
Instabilities in the flow due to shear between the jet and the star or
backflow from the jet head result in fluctuating jet speeds.  Recent
numerical simulations of relativistic jets propagation through
collapsars demonstrate the presence of the variability some with
time scales of $\sim 0.1$ sec (Zhang, Woosley \& MacFadyen).

MacFadyen, Woosley \& Heger, 2001 (hereafter MWH) showed that jets of
equivalent energy injected into a stellar envelope can be focussed by
the pressure of the star.  The degree of focusing is a function of
the assumed entropy of the jet which was parametrized by
$E_{int}/E_{tot}$.  ``Colder'' jets were squeezed into tightly collimated
flow while ``hotter'' jets of the same energy expanded sideways (See
MWH Fig 10).  The result was that ``hotter'' jets are broader and
sweep up a larger mass of stellar material along the rotation axis of
the star.  While the calculations in MWH were non-relativistic, the
results should hold in special relativity since the asymptotic Lorentz
factor is an inverse function of the swept up mass.  Current fully
relativistic numerical simulations indicate the results do hold
(Zhang, Woosley \& MacFadyen).

MWH also experimented with the effect of the collapsar density
structure on jet collimation and found a large difference in jet
opening angle for two density structures corresponding to low and high
disk mass.  The difference was due to the assumed viscosity parameter,
but may also result from initial angular momentum and density structure
of the progenitor stars.  Calculations of the collapsar density
structures and their effect on relativistic jet opening angle are
currently underway.

\section{Typical GRB energy}

In the collapsar model, the energy powering the GRB originates in an
accretion torus surrounding a rotating black hole of several solar
masses.  The stellar progenitor must lose its envelope in order to
have a sufficiently small radius to allow a relativistic jet formed
near the accreting black hole to pierce the star in a GRB time scale
i.e. $R_{*} < 10 sec ~(c/2) \sim 1.5 \times 10^{11}$ cm.  In
addition the range of angular momentum conducive to GRB formation may
be narrow.  It is constrained on the low end by the angular momentum
sufficient to form a disk and on the upper end, so that the disk is
small enough to cool at least partially by neutrino emission (though
these neutrino need not participate directly in jet formation).  Disks
which are inefficiently cooled are convective leading to outflows and
inefficient accretion (MacFadyen \& Woosley 1999; Narayan et. al).  
The successful GRB-producing collapsars may therefore be expected to have
roughly similar energetics.

Another requirement for GRB production is that the jets have sufficient 
momentum flux to overcome the ram pressure of the column of gas
accreting along the stellar rotation axis. Since the momentum flux
decreases with a larger opening angle, there is a 
maximum opening angle which will not be swamped by the accretion ram.
Equivalently for a given opening angle there is a minimum momentum
flux capable of launching a jet.  Given an observed maximum jet opening
angle, this sets a lower limit for energy flux capable of producing a
GRB.  Given a similar typical time scale this implies a similar total
energy.

\section{Internal Shocks and the Temporal Structure}
Internal shocks arise in a relativistic wind with a nonuniform velocity
when the fast moving flow catches up the slower one. The wind can be
modelled by a succession of relativistic shells (Kobayashi, Piran \& Sari
1997, hereafter KPS). A collision of two shells is the elementary
process, and produces a single pulse of gamma-ray. 
Three time scales, the cooling time, the
hydrodynamic time and the angular spreading time, are relevant to the
pulse width. With the relevant parameters the cooling time is negligible
compared to the other two time scales (Sari, Narayan \& Piran 1996). 
Let $d$ and $D$ be the width and separation of the shells. In other
words, we assume that the central engine operates for a period $d/c$ and
then is quiet for a period $D/c$. 
The hydrodynamic time scale $\sim d/c$ and the angular
spreading time scale $\sim D/c$ determine the rise and the decay time
of the pulse, respectively. Since most observed pulses rise more quickly
than they decay, the pulse width is mainly determined by the angular
spreading time $\sim D/c$ (Norris et al. 1996 ; Ryde \& Petrosian 2001).

The whole light curve of a GRB is given by the superposition of the 
resulting pulses from each collision. Since all shells are moving 
towards us with almost speed of light, we observe pulses arising from 
collision between shells mostly according to their positions inside the 
wind, i.e., according to the time when those shells were emitted by the 
central engine (KPS; Nakar \& Piran 2001). The relative positions of the 
shells inside the wind are also determined by the
separations. Therefore, the variability time scales in the temporal 
profile reflect the shell separations at the central engine well.

The observed distribution of peak separations $\{D/c\}$ has a large 
dispersion (Nakar \& Piran 2001). While the fastest variability time
scale in GRBs is about a millisecond, the largest peak separations or 
the total durations are usually very much longer. Long bursts have 
durations from 10 to 1000sec. Clearly, whatever is the physical 
mechanism behind the GRB production, it acts on a much longer timescale 
than the fastest dynamical time. Many GRB models involve accretion onto a
compact object, usually a black hole. If the accretion disk is fed by
fallback of material after a supernova explosion, as in the collapsar
model, then the duration and large peak separations are determined by
fallback, not the dynamical accretion time.  The large dispersion of 
peak separations could be explained in the collapsar model.

It has recently been shown that the Thomson optical depth
due to $e^\pm$ pairs produced by synchrotron photons plays an important
role in the internal shock model (Guetta, Spada \& Waxman 2001; Asano \&
Kobayashi 2002). In order to obtain high radiative efficiency and the 
characteristic clustering of spectral break energies of GRBs in the 
range 0.1-1 MeV, the collision radii are required to be similar to the
photosphere radius. Since collisions producing  narrow pulses occur
at small radii $\propto D$, the photosphere might obscure these, leaving
only the wider pulses visible. This will make the temporal profile
smooth. In a wider jet, the typical  
Lorentz factor $\Gamma$ is smaller, a larger fraction of the collisions 
occur at small radii $\propto \Gamma^2$ below the photosphere,
therefore, the smoothing effect is expected to be stronger. This might
explain the correlation between luminosity and variability in GRBs.

\section{Shell Model}

We discuss the smoothing effect by using a multiple-shell model. 
We represent the irregular wind by relativistic shells in a manner
similar to that in KPS. Because of the relativistic beaming effect, we
can study the emission from a jet by using spherical shells with an 
isotropic explosion energy $E_{iso}(\theta)=4 E/\theta^2$ where $E$ is
the geometrically corrected explosion energy and $\theta$ is the opening 
angle. The gamma-ray energy released from a GRB is narrowly clustered around
$10^{51}$erg (Frail et al. 2001), and the conversion efficiency 
from the explosion energy into the gamma-rays is about $10\%$ (Guetta,
Spada \& Waxman 2001; Kobayashi \& Sari 2001). Then, $E \sim 10^{52}$ erg. 

\subsection{Two-Shell Collision and Photosphere}

Consider a two shell collision which is the elementary process in the
multiple shell evolution. A rapid shell with Lorentz factor $\Gamma_r$
and mass $m_r$ catches up to a slower one with $\Gamma_s$ and $m_s$, and
the two merge to temporarily form a single shell. Using conservation of
energy and momentum we calculate the Lorentz factor of the merged shell
to be  
\begin{equation}
\Gamma_m \sim \sqrt{\frac{m_r\Gamma_r+m_s\Gamma_s}
{m_r/\Gamma_r+m_s/\Gamma_s}}. 
\end{equation}
The internal energy of the merged shell is the
difference of kinetic energy before and after the collision, 
$E_{int}=m_rc^2(\Gamma_r-\Gamma_m)+m_sc^2(\Gamma_s-\Gamma_m)$. 
We assume that electrons are accelerated to a power-law distribution
of Lorentz factor $\gamma_e$, with a minimum Lorentz factor $\gamma_m$: 
$N(\gamma_e)d\gamma_e \propto \gamma_e^{-p}d\gamma_e, \gamma_e \ge
\gamma_m$. Throughout this paper, we use the standard choice $p=2.2$.
A constant fraction $\epsilon_B$ of the internal energy goes into 
magnetic energy. The energy distributed to the electrons is radiated 
via synchrotron emission.  

A significant fraction of the kinetic energy could be converted to the 
internal energy if the relative velocity of the two shells is
relativistic. However, it is not reasonable that all the internal energy 
is emitted, because electrons do not have most of the internal energy, 
and protons do not cool. Defining $\epsilon_e$ as the fraction of energy 
given to electrons, we expect $\epsilon_e <1$. Even at equipartition 
with protons $\epsilon_e$ is $1/2$

If $\epsilon_e <1$, the merger produced by a collision is expected
to stay hot after the emission. As a result, the merger
will spread, transforming the remaining internal energy back to kinetic 
energy. Kobayashi and Sari(2001) studied such a process by using a
hydrodynamic code to show that the spreading leads to a formation of
two shell structures, rather than a homogeneous wide shell. This is a 
planar analogy of the evolution of a relativistic fireball (Kobayashi, 
Piran and Sari 1999). 

Let us consider a two shell collision in the rapid shell rest frame.
When the slow shell begins to interact with the rapid shell material, 
two shocks are formed: a forward shock propagating into the rapid shell 
and a reverse shock propagating into the slow shell. After the reverse 
shock crosses the slow shell, the profile of the shocked rapid shell 
material approaches to that of its fireball analogy: the ``blast wave'' 
which sweeps and collects the rapid shell material. Once the shock wave 
crosses the rapid shell, the hydrodynamical structure is as follows. 
There is a shocked rapid shell,  the analogy of the ``blast wave'' and 
the shocked slow shell, which is the analogy of the ``fireball ejecta''.  

When the evolution of multiple shells is consider in the following 
sections, shell collisions happen around a radius $\sim \Gamma_s^2 D$,
and therefore the time interval of the collisions is order of 
$\sim \Gamma_s^2D/c$. Since the hydrodynamic time scale (shock crossing 
time) $\sim \Gamma_s^2 d/c$ is much shorter than the interval, each two 
shell interaction is completed in a moment, and no shell collides into 
the merger during the interaction. A simplified description of
the spreading effect is to assume that the two shells reflects with a 
smaller relative velocity (Kobayashi \& Sari 2001).

The synchrotron photons can be scattered by electrons within the 
merger. The Thomson optical depth is increased significantly when
taking into account $e^\pm$ pairs produced by internal shocks
(Guetta, Spada \& Waxman 2001; Asano \& Kobayashi 2001).
The typical energy of the synchrotron emission, in the shell frame, is
\begin{equation}
h\nu_m^\prime \sim 10 ~\mbox{keV} ~\epsilon_e^2 \epsilon_B^{1/2}
E_{int,52}^{1/2}R_{14}^{-1}d_{7}^{-1/2}\Gamma_{m,2}^{-1},
\end{equation}
where $f=10^x f_x$, $E_{int}$ is in units of erg, the collision radius
$R$ and the shell width $d$ are in units of cm. This energy is well
below the pair production threshold. However, since the photon spectrum
extends to high energy as a power law  
\begin{equation}
F_{\nu^\prime} \propto \left\{
              \begin{array}{@{\,}ll}
              \nu^{\prime -1/2}& \nu^\prime< \nu^\prime_m \\
              \nu^{\prime -p/2}& \nu^\prime> \nu^\prime_m 
\end{array} 
\right.
\end{equation}
there exists a large number of photons beyond the threshold 
$h\nu^\prime \sim m_e c^2$. The pairs produced by these photons can
contribute significantly to the Thomson optical depth. 
In order to take the effect of pair production into account, we
determine for each collision the typical photon energy $h\nu^\prime_m$ 
and the amount of energy emitted above the threshold.
\begin{equation}
E^\prime_{\pm}\sim  \paren{\frac{h\nu^\prime_m}{m_ec^2}}^{-p/2+1}
\frac{\epsilon_e E_{int}}{(p-1)\Gamma_m}
\end{equation}
Since the number density of the pairs is given by 
$n^\prime_\pm \sim E^\prime_{\pm}/4\pi m_ec^2R^2d\Gamma_m$, the 
photosphere radius $R_{\pm}$, where the Thomson optical depth
becomes unity, is given by
\begin{equation}
R_\pm \sim \paren{\frac{\sigma_T E^\prime_\pm}{4\pi m_ec^2}}^{1/2}.
\label{eq:rph}
\end{equation}
If a collision happens below the photosphere, the whole internal energy
produced by the collision is converted to kinetic energy again via the
shell spreading. We assume that the two shells just reflect with the same 
relative velocity. 

\subsection{Four Shell Evolution: an example case}

\begin{figure}
\plotone{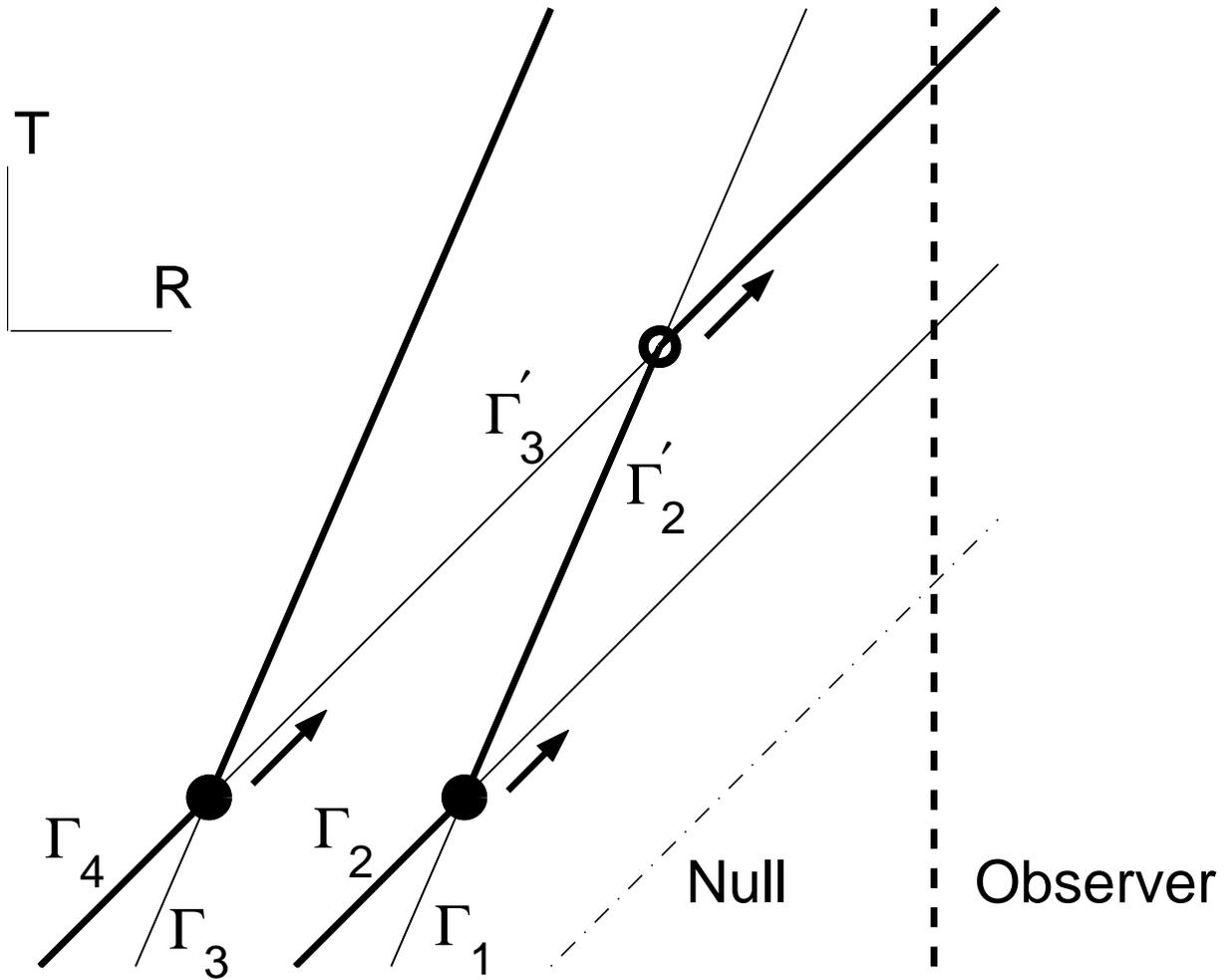}
 \caption{ Emission from collisions. Trajectories: shells ejected from 
the central engine with hight Lorentz factor (thick solid), ones ejected 
with low Lorentz factor (thin solid) and  observer (dashed). 
Collisions (circles) produce pulses of photons (arrows).
The dashed dotted line shows the null geodesic passing through the
central engine at $T=0$.
 \label{fig:trd}} 
 \end{figure}

We here consider a evolution of four shells which is a basic component 
in multiple shell interaction. Since the observations require that 
shell widths should be much smaller than the separations, we assume 
infinitesimally thin shells for simplicity. We assign an index $i$, 
($i=1,4$) to each shell according to the order of 
the emission from the central engine (see figure \ref{fig:trd}). 
The central engine ejects each shell with a Lorentz factor 
$\Gamma_i \gg 1$ at time $T_i$. We choose the origin of time such 
that the first shell is emitted at $T_1=0$. For convenience 
we define $D_i\equiv c\paren{T_i-T_{i-1}}$ as the separation between 
shell $i-1$ and shell $i$. 

To evaluate the conversion efficiency or the temporal properties of 
the internal shock process, it is important to understand how high
velocity shells interact with slower ones and dissipate kinetic energy.
Thus, we consider a case that the Lorentz factors of the shell 2 
and the shell 4 are much higher than those of the shell 1 and the shell 3
$\Gamma_2, \Gamma_4 \gg \Gamma_1, \Gamma_3$.
The collision between the shell 1 and the shell 2 takes place at time 
$T \sim T_2+2\Gamma_1^2 D_2/c$ and radius $R \sim 2\Gamma_1^2 D_2$.
The gamma-ray pulse produced by this collision arrives at the observer
at the observer time $t=T-R/c \sim T_2+o(\Gamma_1^2/\Gamma_2^2)$. We 
chose the origin of the observer time such that if a pulse was emitted 
from the engine at $T=0$, it would arrive at the observer at $t=0$ 
(dashed dotted line in figure \ref{fig:trd}). 
The pulse produced by the collision between the shell 3 and the shell 
4 comes to the observer at $t \sim T_4+o(\Gamma_3^2/\Gamma_4^2)$.

The mergers (filled circles) spread and cause
a subsequent collision (open circle). Using the Lorentz factor 
of the shell 2 and the shell 3 after the collisions $\Gamma_2^\prime$ 
and $\Gamma_3^\prime$, the arrival time of the pulse from the
subsequent collision can be estimated as 
$t \sim T_4+o(\Gamma_2^{\prime2}/\Gamma_3^{\prime2})$.
If $\Gamma_2^\prime/\Gamma_3^\prime\ll 1$, the pulses from the 
collisions between the shell 3 and the shell 4 and between the shell 2
and the shell 3 are observed almost at the same time. If 
$\Gamma_2^\prime \sim \Gamma_3^\prime$, the collision can produce only
a weak and soft pulse, it does not contribute to the main feature
of the GRB light curve. Therefore, we observe pulses arising from the 
collisions mostly according to the time when high velocity shells were 
emitted by the central engine.

\subsection{Multiple-Shells}

We consider a wind consisting of $N$ shells. Each shell is characterised
by four variables: Lorentz factor $\Gamma_i$, mass $m_i$, width $d_i$
and the distance to the outer neighbour shell $D_i$.  As we wrote in
section 4, most observed pulses in GRB light curves rise more quickly
than decay, this means that the widths are smaller than the
separations. Since, in such a case, light curves are not sensitive to 
the distribution of widths, for simplicity the width is assumed to be
constant $d=D_{min}$. Though we will show numerical results with various
distributions of $\Gamma_i$, $m_i$ and $D_i$ later, we first consider 
simple distributions with which we can give simple arguments to 
understand the numerical results. We assume that the Lorentz factors and 
the separations are distributed uniformly in logarithmic spaces; between
$\Gamma_{min}$ and $\Gamma_{max}$ and between $D_{min}$ and $D_{max}$,
respectively. We assume that the shells have equal mass. The internal
shock process becomes very efficient if shells have the same mass and
if the dispersion of the initial Lorentz factors is large (Beloborodov
2000; Kobayashi and Sari 2001). The value of mass is normalised by
using the isotropic explosion energy $E_{iso}(\theta)$ as
$m=E_{iso}/\Sigma\Gamma_ic^2$.    

Our key assumption is a correlation between the Lorentz factors and the
opening angle: wider jets have lower Lorentz factors  
$\Gamma \propto \theta^{-\alpha}$. 
The observed wide jets with $\theta \sim 0.2$ also should have
ultra-relativistic Lorentz factors to be optically thin to high energy
photons, we assume $\Gamma_{max}(\theta)=\Gamma_0 ~(\theta/0.2)^{-\alpha}$
and $\Gamma_0=100$

The evolution of the system in time is basically the same as in KPS97,
but the photospheric effect is considered. We follow the evolution of
shells until there are no more collisions, i.e. until the shells are 
ordered by increasing values of the Lorentz factors.
The numerical light curves in the 50 - 300 keV BATSE band are plotted in
Figure \ref{fig:profiles}a-c for a model with different opening angles
$\theta=$ 0.2, 0.06 or 0.02. Since we have assumed $\alpha=1$, 
these correspond to $\Gamma_{max} = 100, 333$ and $1000$, respectively.
$N=100, \Gamma_{max}/\Gamma_{min}=10, D_{min}/c=1$msec, $D_{max}/c=1$sec
and $\epsilon_e=\epsilon_B=0.1$ were also assumed. As we expected, 
the temporal profiles are more variable for narrower jets with higher
Lorentz factors. 

The definition of variability by Fenimore \& Ramirez-Ruiz (2000) and
that by Reichart et al (2001) are slightly different, but both relate 
the variability to the square of the time history after removing low
frequencies by smoothing. To evaluate the variability of numerical
light curves, we use a simplified version of the Reichart et al. (2001)
variability,
\begin{equation}
V=\frac{\Sigma \paren{C_i-\angle{C}}^2}{\Sigma C_i^2}
\label{eq:var}
\end{equation}
where $C_i$ is 50-300 keV photon counts at time $t_i$ with 64 msec resolution 
(the BATSE resolution), and $\angle{C}$ is the counts smoothed with a
boxcar window with a time scale equal to the smallest fraction of the
burst time history that contains a fraction $f$ of the total counts. 
Reichart et al. (2001) found that $f=45\%$ gives a robust definition of
variability. Using the formula (\ref{eq:var}) with $f=0.45$, we
evaluated the variability measures as 
$V=0.06, 0.10$ and $0.25$ for Figure \ref{fig:profiles}a,
\ref{fig:profiles}b and \ref{fig:profiles}c, respectively.  

Reichart et al (2001) found that the isotropic peak luminosities
$L$ are correlated with the variability measures,
$L\propto V^{3.3^{+1.1}_{-0.9}}$. We now show that such
a correlation exists in our numerical simulations also. A numerical
temporal profile depends on the specific realizations of random
variables that are assigned to each shell, e.g. random Lorentz factors 
and random separations, as well as on the model parameters. For a given
$\theta$ and the same model parameters as in Figure \ref{fig:profiles}, 
we calculate the
temporal profiles for 100 realizations, and evaluate the mean isotropic
peak luminosity and the mean variability measure. Here the peak
luminosity is calculated with 1 sec resolution as Reichart et al (2001)
had analysed. In Figure \ref{fig:var_L}a, the thick solid line shows the mean
values, while the thin solid lines depict the 1$\sigma$ error.  We also
plot the opening  angle-variability relation in Figure
\ref{fig:var_L}b.  The numerical results reasonably fit the
observational data, and can be understood using the following arguments.

\subsection{Analytic Estimate}

Numerous collisions happen during an evolution of multiple-shells. 
Each collision produces a pulse. However, the main pulses are produced
by collisions between the fastest shells $\sim \Gamma_{max}$ 
and the slowest shells $\sim \Gamma_{min}$. Since such a
collision happens at $R \sim \Gamma_{min}^2 D_i$ for an initial
separation $D_i$, we can define two characteristic collision
radii, $\Gamma_{min}^2D_{min}$ and $\Gamma_{min}^2D_{max}$.
In this equal mass case, mass is given by
$mc^2\sim 2E_{iso}/N\Gamma_{max}$. The internal energy produced
by a main collision and the Lorentz factor of the merger are 
$E_{int} \sim E_{iso}/N$ and
$\Gamma_m\sim\sqrt{\Gamma_{max}\Gamma_{min}}$, respectively. The Thomson
photosphere radius due to $e^\pm$ pairs from pair production interaction of
synchrotron photons can be estimated by assuming that a significant
fraction $\sim 10\%$ of the radiative energy  $\epsilon_e E_{int}$ is 
converted to pairs.
\begin{eqnarray}
R_\pm &\sim& \sqrt{\frac{\sigma_T \epsilon_e E_{iso}}
{40\pi m_ec^2N\Gamma_{max}^{1/2}\Gamma_{min}^{1/2}}} \label{eq:photos} \\
      &\sim& 5 \times 10^{14}~\mbox{cm}  ~E_{52}^{1/2}N_2^{-1/2}
\Gamma_{0,2}^{-1/2}(\theta/0.2)^{-1/2},
\end{eqnarray}
hereafter $\epsilon_e=0.1$ is assumed.

If $R_\pm > \Gamma_{min}^2D_{max}$, the first collisions for most shells
occur below the photosphere. However, since the shells reflect each
other with the same velocity, it is still possible to produce bright pulses
when the shells propagate beyond the photosphere and collide into other
shells. If an enormous number of collisions and reflections
happen below the photosphere, the shells are ordered with increasing
values of the Lorentz factors and no collision happens any
more. In order to produce bright pulses, the shells should
come out from the photosphere before the slowest shells are sorted out
at the tail of the wind with a total width  $\Delta \sim ND_{max}/2$.

The equalities, $R_\pm=\Gamma_{min}^2D_{min}$ and
$R_\pm=\Gamma_{min}^2\Delta$ give two opening angles.
\begin{eqnarray}
\theta_1 &\sim& 10^{-4} ~E_{52}^{-1/3}N_2^{1/3}\Gamma_{0,2}^{5/3}
D_{min,-3}^{2/3}\\
\theta_2 &\sim& 10^{-1} ~E_{52}^{-1/3}N_2\Gamma_{0,2}^{5/3}
D_{max,0}^{2/3}
\end{eqnarray}
where $D_{min,-3}=(D_{min}/c)/10^{-3}$sec and $D_{max,0}=(D_{max}/c)/1$sec.
Using these opening angles, GRBs are classified
into the following three cases. 
(1) Narrow Jet Case: 
$\theta < \theta_1$. Since all collisions occur above the photosphere,
the variability should be less dependent on $\theta$ and only on the 
distribution of the separations at the central engine.  
(2) Intermediate Jet Case: $\theta_1 < \theta < \theta_2$. 
The variability of the temporal profile should highly depend on $\theta$. 
Some of the main
collisions occur above the photosphere to produce bright pulses
wider than 
\begin{equation}
\delta t\sim R_\pm/2c\Gamma_{min}^2
\sim 10^{-1} \mbox{sec} ~E_{52}^{1/2}N_2^{-1/2}\Gamma_{0,2}^{-5/2}
(\theta/10^{-2})^{3/2}. 
\label{eq:angtime}
\end{equation}
(3) Wide Jet Case: $\theta > \theta_2$. The slowest shells are sorted
out at the tail of the wind within the photosphere, only minor
collisions happen. The resulting bursts become dimer.  

Figure \ref{fig:profiles}a is an example of the Wide Jet Case, 
while figure \ref{fig:profiles}b and
\ref{fig:profiles}c correspond to the Intermediate Jet Case. Equation
(\ref{eq:angtime}) gives rough estimates on the widths of main pulses
for these Intermediate Jet Cases, $\delta t\sim 3$ sec $(\theta=0.06)$ and 
$\sim 0.5$ sec ($\theta=0.02$). Main pulses are actually wider in figure 
\ref{fig:profiles}b. From figure \ref{fig:var_L}b, the variability 
(thick solid line) is about $0.08$ at $\theta=\theta_2\sim0.1$. In 
figure \ref{fig:var_L}a, a linear fit to the numerical data for 
$V>0.08$ is $L \propto V^{3.8}$,
which is consistent with a power law reported 
by Fenimore\&Ramirez-Ruiz(2000) and by Reichart et al. (2001).
Since the shells are ordered by velocity within the photosphere, 
the numerical luminosity rapidly decreases for $V<0.08$.

\section{Other Initial Conditions}

In our numerical simulation, we have assumed that a wider jet has
a lower Lorentz factor $\Gamma \propto \theta^{-\alpha}$. 
A second assumption is that the shells have equal
mass. A third concerns the distributions of the initial separations and
the initial Lorentz factors. In this section, we discuss how
numerical results depend on these assumptions.

Since the dynamics of the blast wave is determined only by two
parameters: isotropic explosion energy and ambient matter density, 
afterglow emission does not depend on the initial Lorentz factor.
The current afterglow observations detect radiation from several 
hours after the bursts. At this stage, the Lorentz factor of the 
blast wave is less than $\sim 10$. Then, the initial Lorentz factor can 
not be directly constrained through afterglow modelling. However, 
the Lorentz factor at the beginning of the afterglow stage can be 
estimated if one knows when the afterglow began. 

An optical flash was observed during an exceptionally bright burst 
GRB990123, by using this peak time, the Lorentz factor was estimated 
as about 300 (Sari \& Piran 1999; Kobayashi 2000). However, the optical 
flash was detected only for it so far. Panaitescu 
and Kumar (2002) assumed that the observed GRB duration is a good 
measure of the peak time, and
found a correlation $\Gamma \propto \theta^{-0.3}$ within ten events for
which opening angles are well constrained from the afterglow break
time. Salmonson and Galama (2002) also have recently suggested the
anticorrelation of $\Gamma$ with $\theta$, based on the positive
correlation they observed in several cases between the GRB pulse
lag time and the afterglow break time. However, the dependence they
infer is much stronger $\Gamma \propto \theta^{-8/3}$. Therefore, we
have done simulations with $\alpha=1/2$ and 2 also. 

The numerical results with $\alpha=2$ are shown in figure 
\ref{fig:var_L}a and \ref{fig:var_L}b 
(dashed lines). The other parameters are the same with the case of
$\alpha=1$ (solid line). For a larger value of $\alpha$, the Lorentz
factor depends more strongly on the opening angle, while the photosphere
radius becomes insensitive $R_\pm \propto \theta^{(\alpha-2)/2}$. Since
the variability $V$ increases more rapidly with decreasing of the
opening angle, the variability - luminosity relation becomes gentle. For
$\alpha=1/2$, the numerical variability is insensitive to the opening
angle. When we calculate $V$ in a range of $0.02 < \theta < 0.5$, it is
almost constant $V\sim 0.05$. Therefore, the variability - luminosity
relation becomes very steep. The numerical results with $\alpha\sim1$
give the best fit to the observations. The mass loading at the central
engine is $M \sim E/c^2\Gamma \propto \theta$, which is in a good
agreement with a correlation $M \propto \theta^{1.1}$ reported by
Panaitescu and Kumar (2002).

We next discuss the assumption about mass. We also studied the cases
that the shells have the same energy initially or have random masses.
The random masses are assumed to be uniformly distributed between $m$
and $10 m$. The value of $m$ is normalised at the beginning of the 
computation by the isotropic energy $E_{iso}=\Sigma m_ic^2\Gamma_i$. 
Under these mass assumptions, we calculate the temporal profiles for 
100 realizations, and evaluate the mean isotropic peak luminosity and
the mean variability measure. Figure \ref{fig:var_L2}a 
and \ref{fig:var_L2}b present the
variability - luminosity and variability - opening angle relations.
The equal energy (dashed line) and random mass (dashed
dotted line) cases give similar results to the equal mass case (solid line).
For a given $\theta$, the variability in the random mass case is 
larger than in the equal mass case. However, by choosing a smaller 
normalization of Lorentz factor $\Gamma_0 \sim 80$, we can shift the
line horizontally, and can get a more similar relation. Therefore, the 
equal mass assumption is not essential to get the observed correlation.

GRB light curves contain many pulses with widths of msec to sec time
scales. Since widths are determined by the initial separations,
we considered a simple case that the initial separations $D_i$ are
uniformly distributed in logarithmic spaces between $1$ msec and
$1$sec. However, it is known that pulse widths have a lognormal 
distribution (Nakar \& Piran 2001). We here consider a
lognormal random intervals: a mean $\langle\log_{10}(D)\rangle=-2$ and 
a dispersion $\delta(\log_{10}(D))=1$. The thick solid line in figure 
\ref{fig:var_L3} 
shows the numerical results, and is similar to the case of uniform 
distribution (thin solid line).

In order to achieve a high conversion efficiency $(>\mbox{a few} \%)$
from the explosion energy to gamma-ray in the internal shock process 
$\Gamma_{max}/\Gamma_{min}$ should be larger than $\sim 10$ (Kobayashi
\& Sari 2001).  The dashed dotted line in figure \ref{fig:var_L3} 
 depicts the case of
$\Gamma_{max}/\Gamma_{min}=100$, which steepness is comparable to the
case of $\Gamma_{max}/\Gamma_{min}=10$ (thin solid). For a given
luminosity, the variability is larger in this case. However, by using a
smaller $\Gamma_0$, we can get a better fit to the observations. 
We also have done numerical simulations with a Gaussian random Lorentz
factors where a mean value $\langle\Gamma\rangle = 50 (\theta/0.2)^{-1}$
and a dispersion $ \delta \Gamma=\langle\Gamma\rangle$ are assumed. We
truncated the initial Lorentz factor distribution for $\Gamma < 2$,
e.g. when we assign a random value to each shell at the beginning, random 
Lorentz factors are generated until $\Gamma_i > 2$ is satisfied. The
numerical results (dashed line) are also similar to the other cases (see
figure \ref{fig:var_L3}). 

We have shown that the variability - luminosity relation exists even
if we make different assumptions on the distributions of masses,
intervals and Lorentz factors
and that the index and the normalization of the variability - luminosity
relation highly depends on $\alpha$ and $\Gamma_0$. Our numerical
results suggest $\alpha \sim 1$ and $\Gamma_0 \sim 10^2$.

Another assumption concerns the values of $\epsilon_e$ and
$\epsilon_B$, fractions of internal energy distributed to electrons and
magnetic field at collisions. We have done another set of simulations with
$\epsilon_e=\epsilon_B=0.3$ and get very similar results.

\section{Conclusions}

We have shown that there exists a correlation between the jet opening
angle $\theta$ and the gamma-ray light curve variability $V$. Though the
correlation is based on only seven events at present and needs to be
further confirmed with more events, it is naturally expected if the
luminosity $L$ is correlated with the variability (Fenimore \&
Ramirez-Ruiz 2000; Reichart et al. 2001), and if GRBs have a standard
energy output (Frail et al. 2001; Panaitescu \& Kumar 2002; Piran et
al. 2001). This correlation might give us a way to measure the opening
angle for a long burst directly from the GRB light curve.  

We have shown that the opening angle - variability relation, or
equivalently, due to the constancy of burst energy, the luminosity
- variability  relation can be interpreted as the correlation between 
the opening angle of a fireball jet and the Lorentz factor. Larger 
opening angles are expected to be associated with greater mass loading 
at the central engine and might result in lower Lorentz factors. We also 
show that such a correlation can be a natural consequence of the
collapsar model. Using a multiple-shell model, we numerically calculate 
the temporal profiles and estimate the luminosity and the variability. 
Our numerical results suggest $\Gamma \propto \theta^{-1}$ or 
equivalently $M \propto \theta$.

If the opening angle of a jet is very wide $\theta \gg \theta_2$,
the shells are almost ordered by increasing values of the Lorentz factors 
in the photosphere, only minor collisions or, at the extreme, only 
external shocks happen.  The resulting dim burst should be smooth and 
has a soft spectrum. GRB980425 and recently reported Fast X-ray Transients 
(Heise et al. 2001; Kippen et al 2001) could be classified into this 
``Wide Jet Case''.

Norris, Marani \& Bonnell (2000) found that the luminosity 
$L$ is inversely proportional to the GRB pulse lag time.
The detailed study on the lag time is beyond the scope of
this paper. However, if the time lag is also determined by the angular 
spreading time scale, which is the key time scale in our model, we 
can get the same relation. In our model, luminosity is scaled by the 
opening angle $L \propto \theta^{-2}$, while the angular 
spreading time is $t_{ang}\propto \Gamma^{-2} \propto \theta^2$. 
Then, we obtain $L \propto t_{ang}^{-1}$. 

\begin{figure}
\plotone{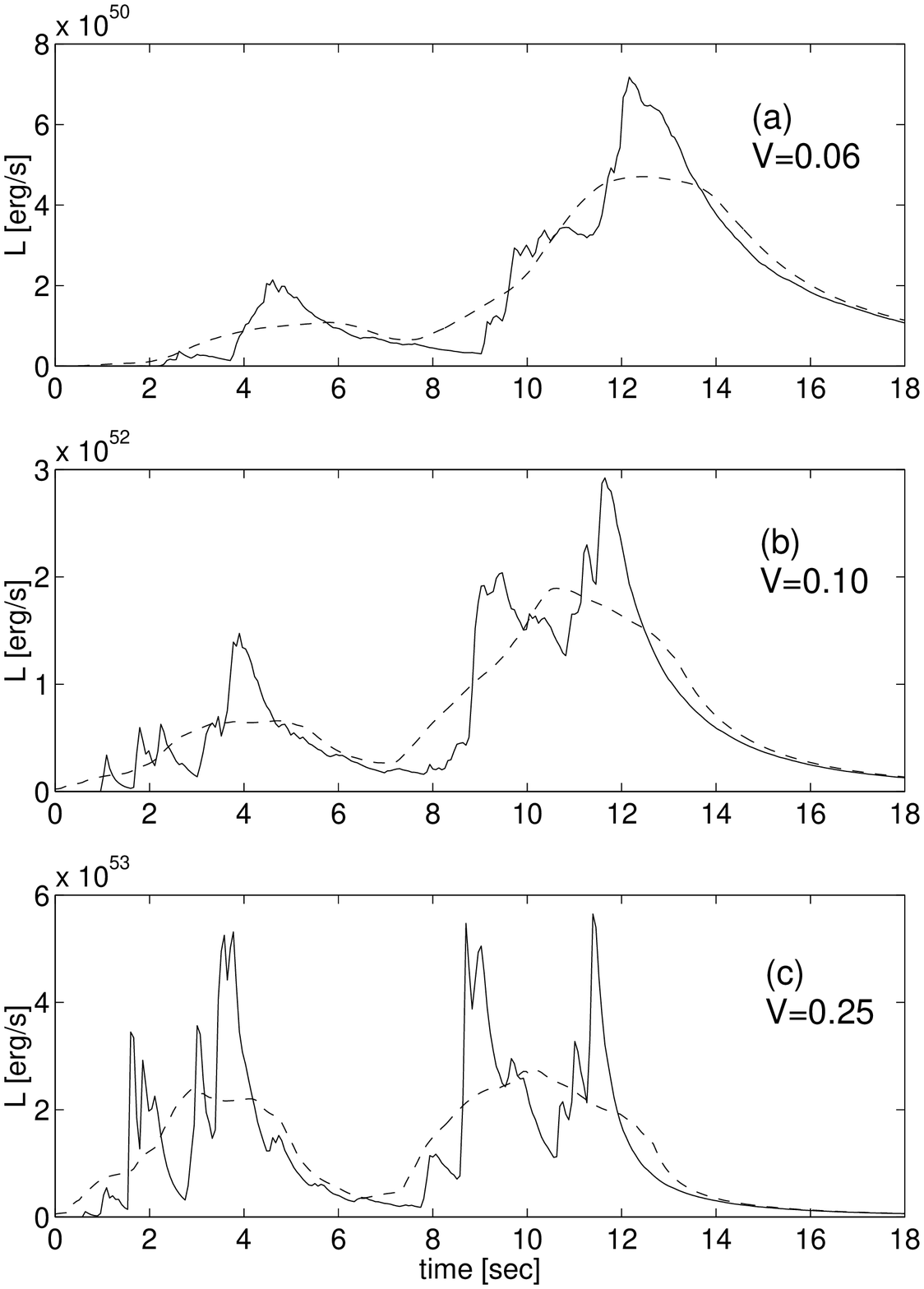}
 \caption{
Temporal structures (50-300 keV) for different opening angles.
(a) $\theta=0.2$. 
(b) $\theta=0.06$.
(c) $\theta=0.02$.
The dashed lines show the structures smoothed with the boxcar
wind. 
 \label{fig:profiles}} 
 \end{figure}

\begin{figure}
\plotone{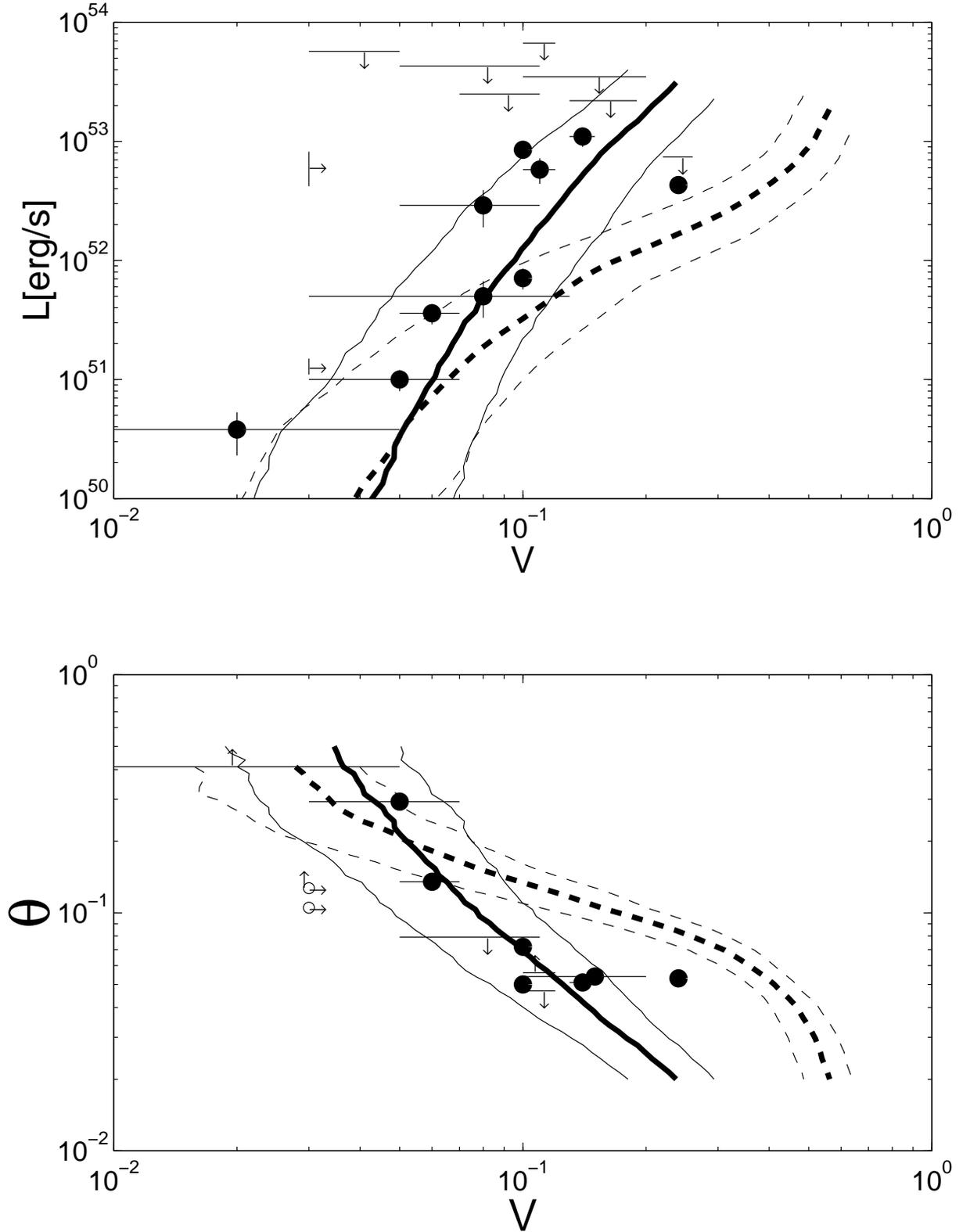}
 \caption{(a) Peak Luminosity  vs.  Variability
with 1 $\sigma$ error bars of 100 random simulations. 
Observational data: measurements (filled circles) and upper or lower limits 
(arrows) are shown from Reichart et al (2001) and Frail et al (2001).
$\alpha=1$(solid line), $\alpha=2$(dashed line)
(b) Opening angles vs. Variability 
 \label{fig:var_L}} 
 \end{figure}

\begin{figure}
\plotone{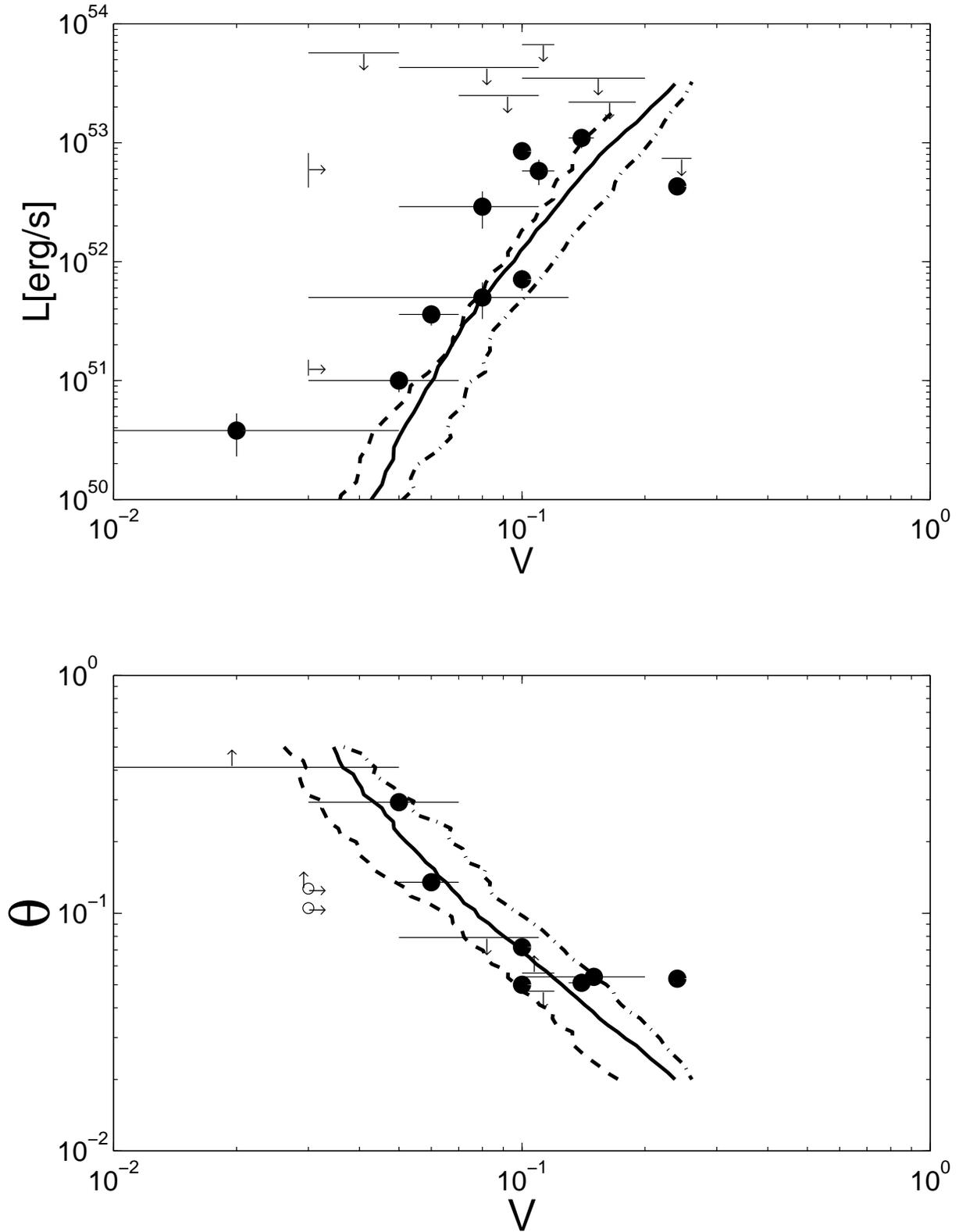}
 \caption{(a) Peak Luminosity vs. Variability : 
mean of 100 random simulations. Equal mass (solid line),
equal energy (dashed line) and random mass (dashed dotted line).
The parameters are the same with Figure \ref{fig:var_L}($\alpha=1$) 
except the mass conditions.
(b) Opening angles vs. Variability 
 \label{fig:var_L2}} 
 \end{figure}

\begin{figure}
\plotone{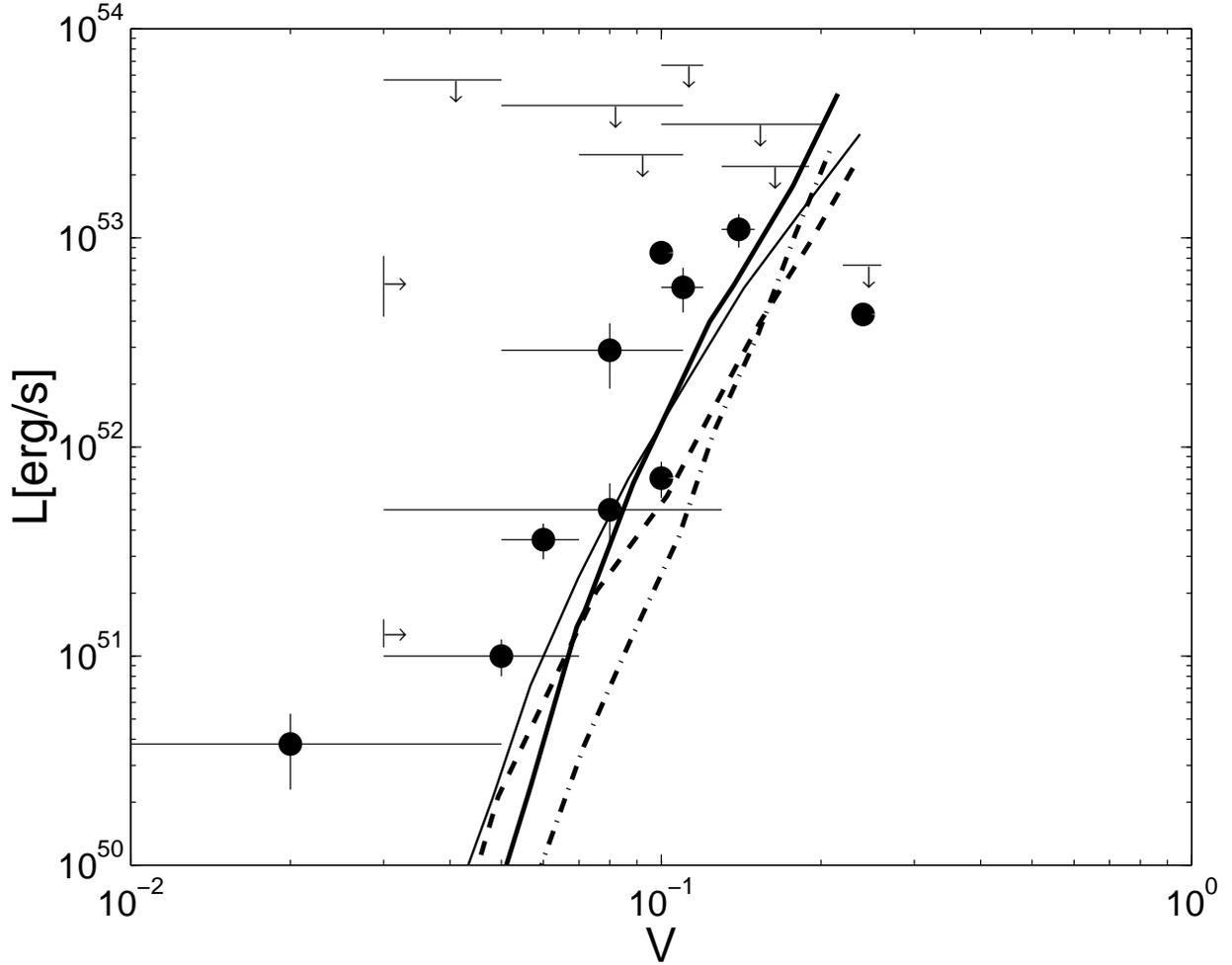}
 \caption{(a) Peak Luminosity vs. Variability : 
mean of 100 random simulations. 
Lognormal random intervals (thick solid).
$\Gamma_{max}/\Gamma_{min}=100$ (dashed dotted) and
Gaussian random Lorentz factors (dashed).
$\Gamma_{max}/\Gamma_{min}=10$, uniform
distributions of intervals and Lorentz factors are assume if 
the value or the distributions are not specified. $\alpha=1$ and equal
mass are assumed in all cases. The thin solid line
depicts the standard case.
 \label{fig:var_L3}} 
 \end{figure}

\begin{figure}
\plotone{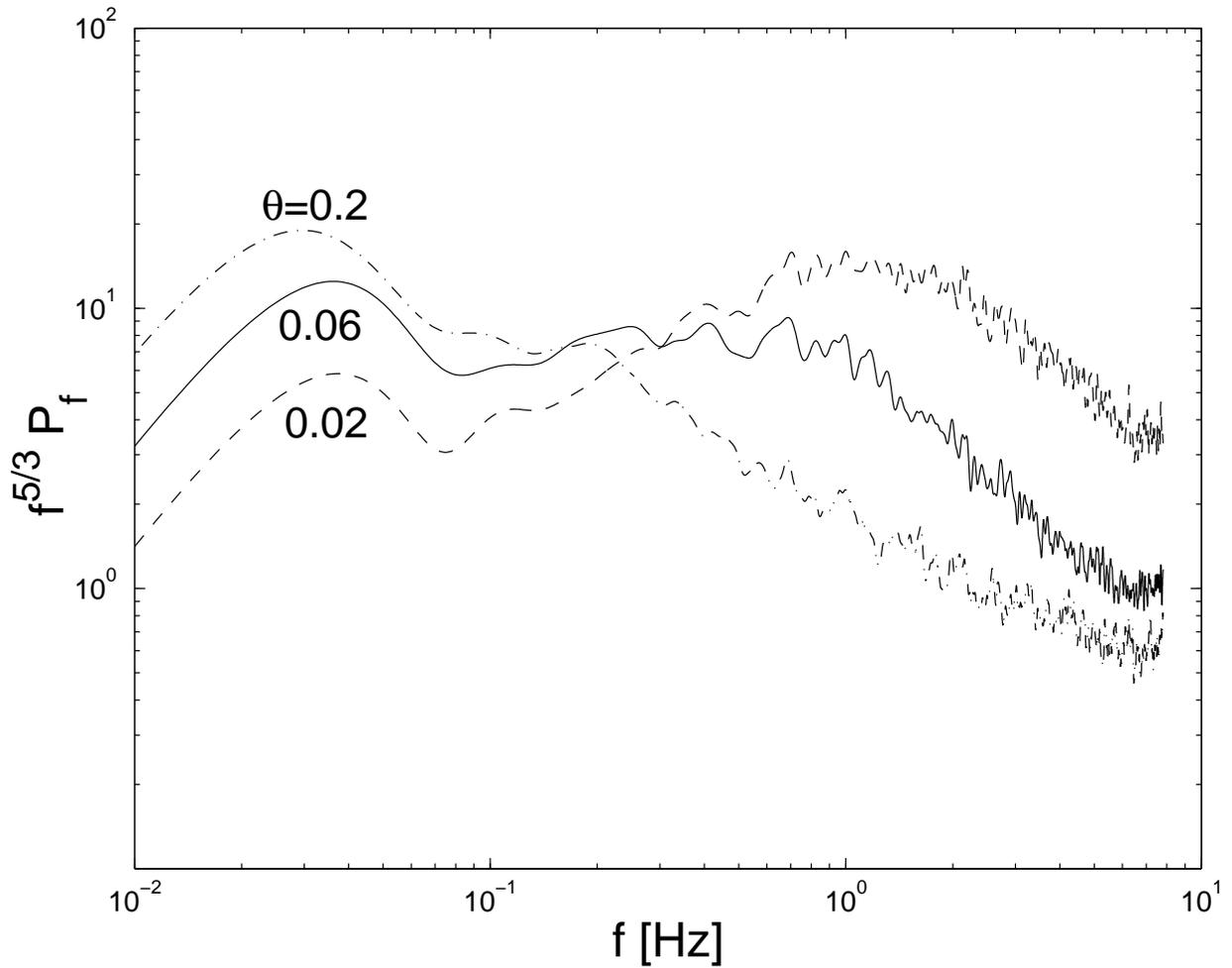}
 \caption{ The averaged PDS for 100 random realizations.
opening angle
$\theta=0.2$(dashed dotted), 0.06 (solid) and 0.02 (dashed).
The initial parameters are the same as in Fig \ref{fig:profiles}
 \label{fig:pds}} 
 \end{figure}

Another measure of variability in light curves is the power density spectra
(PDSs). The observed PDSs show a strong correlation
with the luminosity (Beloborodov, Stern and Svensson 2000). In each 
case of figure \ref{fig:profiles} a, b and c, we calculate the light 
curves for 100 realizations, and evaluate the average PDS $P_f$. 
The averaging means that we sum up the PDSs of the numerical profiles 
with the peak normalisation, and divided by the number of realizations, 
100. The numerical PDSs show a strong correlation with the opening
angles : brighter bursts with smaller opening angles have more variable 
power at high frequency (see figure \ref{fig:pds}). It resembles the 
results for different brightness classes of real bursts (figure 8 in 
Beloborodov, Stern and Svensson 2000).  The numerical results reproduce 
the slope index of $\sim -5/3$ and the 1 Hz break also which Beloborodov,
Stern and Svensson (2000) found in observed light curves.  
The photosphere radius $R_\pm$ is about 
$5\times 10^{14}$ cm and weakly depends on parameters. The collisions 
which produce 1sec pulses happen around a radius 
$\sim 6\times10^{14} (\Gamma_{min}/100)^{2}$ cm. If $\Gamma_{min}\sim
100$, the 1Hz break can be explained by the phtospheric effect. The 
constraints on the activity of the central engine from the PDS study 
will be discussed in the subsequent paper. 

In our model, more variable bursts have higher Lorentz factors, and
therefore are expected to have  higher spectral peak energy. We can 
actually see such a trend in the observational data when correcting
the cosmological expansion effect(Table 1 of Fenimore and Ramirez-Ruiz 
2000). Ramirez-Ruiz and Lloyd-Ronning (2002) recently extended our 
study to examine the spectral property of GRBs. Their numerical 
results reproduce the trend.

Most studies on GRBs assume a jet with a well-defined angle, though this
angle differs for different bursts. However, Rossi, Lazzati \& Rees
(2001) and Zhang \& M\'{e}sz\'{a}ros (2001) proposed an 
alternative model where the jet, rather than having a uniform profile
out to some definite cone angle, has a beam pattern. Even in this 
model, the more luminous part of a jet has a higher Lorentz
factor. Then, our results can be applied to explain the variability - 
luminosity relation.

We would like to thank the Aspen Center for Physics for their 
hospitality and for providing a pleasant working environment where 
this work was initiated. We would also like to thank Nicole M. 
Lloyd-Ronning for useful discussions. We thank the anonymous referee for
valuable suggestions that enhanced this paper. S.K. acknowledges support
from the Japan Society for the Promotion of Science. F.R. acknowledges
support from the Swedish Foundation for International Cooperation in
Research and Higher Education (STINT). A.M. acknowledges support from
DOE ASCI (B347885).

\newpage

\vspace{1cm}
\noindent {\bf References}\newline
Asano,K., \& Kobayashi, S. 2002, in preparation. \newline
Beloborodov,A.M. 2000, ApJ, 539, L25.\newline
Beloborodov,A.M., Stern,B.E. \& Svensson,R. 2000, ApJ, 535, 158.\newline
Fenimore,E., \& Ramirez-Ruiz, E. 2000, Submitted to ApJ, 
(astro-ph/0004176).\newline
Frail, D. A., et al. 2001, ApJ, 562, L55.\newline
Guetta,D., Spada,M., \& Waxman E. 2001, ApJ, 557,399.\newline
Heise, J., et al. 2001, Proc. Second Rome Workshop \newline
Kippen, R. M., et al. 2001, Proc. Second Rome Workshop, 
astro-ph/0102277.\newline
Kobayashi,S. 2000, ApJ, 545, 807.\newline
Kobayashi,S., Piran,T. \& Sari,R. 1997, ApJ, 490, 92.\newline
Kobayashi,S., Piran,T. \& Sari,R. 1999, ApJ, 513, 669.\newline
Kobayashi,S., \& Sari,R. 2001, ApJ, 542, 819.\newline
Lloyd-Ronning,N.M., Fryer,C.L \& Ramirez-Ruiz,E. 2001, ApJ in press.\newline
MacFadyen,A. \& Woosley,S. 1999, ApJ, 524, 262.\newline
M\'{e}sz\'{a}ros, P., \& Rees,M.J. 2000 ApJ, 530, 292.\newline
Nakar,E., \& Piran,T. 2001, MNRSA in press (astro-ph/0103210).\newline
Narayan, R., Piran, T. \&Kumar, P., ApJ 557, 949.\newline
Norris, J.P. et al. 1996, ApJ, 459, 393.\newline
Norris, J.P., Marani,G.F., \& Bonnell,J.T. 2000, ApJ, 534, 248.\newline
Panaitescu,P. \& Kumar,P. 2002, ApJ in press (astro-ph/0109124).\newline
Piran,T., 2000, Phys. Rep., 333, 529.\newline
Piran,T., et al. 2001, ApJL in press (astro-ph/0108033).\newline  
Ramirez-Ruiz, E. \& Lloyd-Ronning, N.M. 2002, New Astronomy in press.\newline
Ramirez-Ruiz, E. \& Fenimore,E. 1999, Presentation at the 1999
Huntsville GRB conference.\newline
Rossi,E., Lazzati,D. \& Rees,M.J. 2002, MNRAS in press.\newline
Reichart,D. E., et al. 2001, ApJ, 552, 57.\newline
Ryde,F., \& Petrosian,V. 2002, in preparation.\newline
Salmonson,J.D. \& Galama,T.J. 2002, ApJ in press 
(astro-ph/0112298).\newline
Sari,R., Narayan, R., \& Piran,T. 1996, ApJL, 473, 204.\newline
Sari,R. \& Piran,T. 1999, ApJL, 517, 109.\newline
Stern,B., Poutanen,J. \& Svensson,R. 1999, ApJ, 510, 312.\newline
Woosley,S. 1993, ApJ 405, 273.\newline
Zhang,B. \& M\'{e}sz\'{a}ros,P. 2002, ApJ in press 
(astro-ph/0112118).\newline
Zhang,W, Woosley, S., \& MacFadyen, A. 2002, in preparation.\newline
\newpage


\end{document}